\documentclass{aastex63}

\usepackage{booktabs}

\usepackage{color}

\received{xxx}
\revised{yyy}
\accepted{zzz}

\submitjournal{ApJ}

\begin{document}

\title{Slowly cooling white dwarfs in  NGC 6752}

\author[0000-0002-8004-549X]{Jianxing Chen}
\affiliation{Dipartimento di Fisica e Astronomia ``Augusto Righi'',
  Alma Mater Studiorum Universit\`a di Bologna, via Piero Gobetti
  93/2, I-40129 Bologna, Italy}
\affiliation{INAF-Osservatorio di Astrofisica e Scienze dello Spazio
  di Bologna, Via Piero Gobetti 93/3 I-40129 Bologna, Italy}
\correspondingauthor{F.R.Ferraro}
\email{francesco.ferraro3@unibo.it}
\author[0000-0002-2165-8528]{Francesco R. Ferraro}
\affiliation{Dipartimento di Fisica e Astronomia ``Augusto Righi'',
  Alma Mater Studiorum Universit\`a di Bologna, via Piero Gobetti
  93/2, I-40129 Bologna, Italy}
\affiliation{INAF-Osservatorio di Astrofisica e Scienze dello Spazio
  di Bologna, Via Piero Gobetti 93/3 I-40129 Bologna, Italy}
\author[0000-0002-5038-3914]{Mario Cadelano} \affiliation{Dipartimento
  di Fisica e Astronomia ``Augusto Righi'', Alma Mater Studiorum
  Universit\`a di Bologna, via Piero Gobetti 93/2, I-40129 Bologna,
  Italy} \affiliation{INAF-Osservatorio di Astrofisica e Scienze dello
  Spazio di Bologna, Via Piero Gobetti 93/3 I-40129 Bologna, Italy}
\author[0000-0002-2744-1928]{Maurizio Salaris}
\affiliation{Astrophysics Research Institute, Liverpool John Moores
  University, Liverpool Science Park, IC2 Building, 146 Brownlow Hill,
  Liverpool L3 5RF, UK}
\affiliation{INAF-Osservatorio Astronomico d'Abruzzo, Via Maggini,
  I-64100 Teramo, Italy}
\author[0000-0001-5613-4938]{Barbara Lanzoni}
\affiliation{Dipartimento di Fisica e Astronomia ``Augusto Righi'',
  Alma Mater Studiorum Universit\`a di Bologna, via Piero Gobetti
  93/2, I-40129 Bologna, Italy}
\affiliation{INAF-Osservatorio di Astrofisica e Scienze dello Spazio
  di Bologna, Via Piero Gobetti 93/3 I-40129 Bologna, Italy}
\author[0000-0002-7104-2107]{Cristina Pallanca}
\affiliation{Dipartimento di Fisica e Astronomia ``Augusto Righi'',
  Alma Mater Studiorum Universit\`a di Bologna, via Piero Gobetti
  93/2, I-40129 Bologna, Italy}
\affiliation{INAF-Osservatorio di Astrofisica e Scienze dello Spazio
  di Bologna, Via Piero Gobetti 93/3 I-40129 Bologna, Italy}
\author[0000-0002-7104-2107]{Leandro G. Althaus} \affiliation{Grupo de
  Evolucion Estelar y Pulsaciones, Facultad de Ciencias Astron\'omicas
  y Geofisicas, Universidad Nacional de La Plata, Paseo del Bosque
  s/n, 1900 La Plata, Argentina} \affiliation{Centro Cient\'{\i}fico
  Tecnol\'ogico CONICET La Plata, Consejo Nacional de Investigaciones
  Cient\'{\i}ficas y T\'ecnicas, Calle 8 No. 1467, B1904CMC La Plata,
  Buenos Aires, Argentina}
\author[0000-0001-5870-3735]{Santi Cassisi}
\affiliation{INAF-Osservatorio Astronomico d'Abruzzo, Via Maggini,
  I-64100 Teramo, Italy}
\affiliation{INFN - Sezione di Pisa, Largo Pontecorvo 3,  
I-56127 Pisa, Italy}
\author[0000-0003-4237-4601]{Emanuele Dalessandro}
\affiliation{INAF-Osservatorio di Astrofisica e Scienze dello Spazio
  di Bologna, Via Piero Gobetti 93/3 I-40129 Bologna, Italy}

\begin{abstract}
Recently, a new class of white dwarfs (``slowly cooling WDs'') has
been identified in the globular cluster M13. The cooling time of these
stars is increased by stable thermonuclear hydrogen burning in their
residual envelope. These WDs are thought to be originated by
horizontal branch (HB) stars populating the HB blue tail, which
skipped the asymptotic giant branch phase.  To further explore this
phenomenon, we took advantage of deep photometric data acquired with
the Hubble Space Telescope in the near-ultraviolet and investigate the
bright portion of the WD cooling sequence in NGC 6752, another
Galactic globular cluster with metallicity, age and HB morphology
similar to M13.  The normalized WD luminosity function derived in NGC
6752 turns out to be impressively similar to that observed in M13, in
agreement with the fact that the stellar mass distribution
along the HB of these two systems is almost identical. As in the case
of M13, the comparison with theoretical predictions is consistent with
$\sim 70\%$ of the investigated WDs evolving at slower rates than
standard, purely cooling WDs.  Thanks to its relatively short distance
from Earth, NGC 6752 photometry reaches a luminosity one order of a
magnitude fainter than the case of M13, allowing us to sample a regime
where the cooling time delay, with respect to standard WD models,
reaches $\sim 300$ Myr.  The results presented in this paper provide
new evidence for the existence of slowly cooling WDs and further
support to the scenario proposing a direct causal connection between
this phenomenon and the horizontal branch morphology of the host
stellar cluster.
\end{abstract}


%%%%%%%%%%%%%%%%%%%%%%%%%%%%%%%%%%%%%%%%%%%%%%%%%%%%%%%%%%%%%%%%%%%%%%%
\section{Introduction}
\label{sec:intro}
White Dwarfs (WDs) are the final evolutionary stage of the vast
majority ($\sim 98\%$) of stars in the Universe
\citep{winget+08}. Indeed, all stars with initial mass
below $8 M_\odot$, with a possible extension to $11 M_\odot$, are
expected to end their evolution as WDs
\citep{corsico+19, woosley+15}. Their
study can provide a large amount of information about the physical
properties and the evolutionary mechanisms of their
progenitors. Moreover, WDs are the ideal stellar 
structures to test physical processes occurring under extreme matter density conditions.

WDs are commonly envisaged as the naked cores of the progenitor stars
that, at the end of their life, have lost the envelope.  Their
evolution in time is generally described as a pure cooling process,
during which WDs evolve towards cooler temperatures and fainter
luminosities because essentially they cannot produce energy, either
through nuclear reactions, or by gravitational contraction, and
radiate away the residual thermal energy of their ions.
%and can just cool down at constant radius).
This produces a tight relation between WD luminosities and their
cooling ages, which has been commonly adopted as cosmic chronometer to
constrain the age of several populations of our Galaxy, including the
disk, globular and open clusters  (e.g., \citealt{hansen+07,
  bedin+10,jeffery+16,kilic+17}).

However, recent models \citep{althaus+15} show that even an extraordinary
small amount of residual hydrogen (a few $10^{-4} M_\odot$) left over from 
the previous evolutionary stages is sufficient to allow quiescent
thermonuclear burning. This can provide a non-negligible source of
energy that slows down the cooling rate, especially in the low-mass
($<0.6 M_\odot$) and low-metallicity ($Z<0.001$) regimes, which are
typical of globular clusters (GCs) in the Galactic halo.  In turn, an
increased evolutionary time-scale naturally translates to an increased
number of WDs for any fixed luminosity, with a consequent observable
impact on the WD luminosity function (LF).

The first observational evidence of slowly cooling WDs has been
recently provided \citep[][herefater C21]{chen+21} by the analysis of
\textit{Hubble Space Telescope} (HST) ultra-deep photometric data
acquired in the near-ultraviolet for M3 (NGC~5272) and M13 (NGC~6205).
These two Galactic GCs are considered ``twin'' systems because they
have very similar iron abundance ([Fe/H]$\simeq -1.5$), age ($t\sim
12.5$ Gyr), total mass, and central density (see, e.g., C21). In spite
of such a similarity, C21 discovered a clear excess of WDs in the LF
of M13 with respect to M3, and showed that the detected overabundance
is very well described by assuming that $\sim 70\%$ of the WD
population in M13 is composed of slowly cooling objects, while the
objects in M3 cool down as expected from standard models. This result
is fully consistent also with the different morphology of the
horizontal branch (HB) observed in the two systems
\citep[e.g.,][]{ferraro+97a, dalessandro+13}: The HB of M13 shows a
pronounced extension to the blue, while no blue tail is observed in
M3, where the HB is populated in both the red and the blue side of the
instability strip, with a sizeable population of RR Lyrae stars.  The
different morphology is the direct manifestation of different stellar
mass distributions along the HB, with the stars populating the blue
tail of M13 being less massive than those at the red edge, and those
on M3 HB.  In turn, the stellar mass during the HB phase is the
parameter mainly determining the occurrence (or not) of the so-called
\lq{third dredge-up\rq}, a mixing process taking place during the
Asymptotic Giant Branch (AGB) phase, which is the evolutionary phase
immediately following the HB and preceding the WD stage.  This process
efficiently mixes the material present in the envelope of AGB stars,
bringing most of the hydrogen deeper into the stellar interior, where
it is burned. As a consequence, stars experiencing the third dredge-up
reach the WD stage with not enough residual hydrogen to have an
efficient burning. On the other hand, stars evolving from the blue
side of the HB have envelope masses too small to reach the AGB and
therefore do not experience the third dredge-up, and they reach the WD stage
with a residual hydrogen envelope thick enough to sustain stable
thermonuclear burning, thus delaying their ageing
\citep{althaus+15}. Most of the stars in M13 (those populating the
blue side of the HB) skip the third dredge-up (the so-called
post-early AGB and AGB-manqu\'e stars) ending their lives as slowly
cooling WDs. On the other hand, according to the observed HB
morphology, the HB stars in M3 have large enough masses to reach the
AGB and therefore evolve later as standard WDs. The presence of slowly
cooling WDs in M13 and the lack thereof in M3 explains the observed
numerical excess in the LF of the former.

C21 therefore demonstrated that the presence of slowly cooling WDs can
be recognized from the analysis of the WD LF, and it is causally
connected to the HB morphology. Specifically, slowly cooling WDs are
expected in clusters with moderate metallicity and blue-tail HB. In
this context, here we present the analysis of the bright portion of
the WD cooling sequence of NGC 6752, a Galactic GC with
%a very extended blue HB and [Fe/H]$=-1.5$
%(\citealp{Harris1996AJ....112.1487H}, 2010 edition).  In terms of
metal abundance and HB morphology very similar to those of M13 (see
its main parameters in Table \ref{tab1}).  NGC 6752 and M13 also share
a very low value of the $R_2$ parameter, defined as the ratio between
the number of AGB and that of HB stars: $R_2=N_{\rm AGB}/N_{\rm HB} =
0.06$-0.07 \citep{sandquist+04apj, cho+05, cassisi+14}. This is
consistent with the findings of \citet{gratton+10aa1}, who empirically
showed that the $R_2$ parameter correlates with the HB morphology
(clusters with blue extended HB have lower $R_2$ values), because the
stars located in the bluest portion of the HB skip the AGB
phase.\footnote{Note that \citet[][see also
  \citealp{campbell+17}]{campbell+13} claimed that all stars of second
generation (i.e., with [Na/Fe]$< 0.2$; see \citealt{gratton+12} for a
review) skip the AGB. However, this result has been challenged by
synthetic HB simulations \citep{cassisi+14}, and by both spectroscopic
\citep{lapenna+16, mucciarelli+19} and photometric observations
\citep{gruyters+17} that found evidence of both first and second
generation stars along the AGB of NGC 6752.} The simulations of
\citet{cassisi+14} also confirmed that the parameter $R_2$ of NGC 6752
is well reproduced by standard HB models, and they show that the
distribution of stellar masses along the HB in this cluster is
essentially the same as in M13 \citep{dalessandro+13}.

Because of its similarity with M13 in terms of metallicity
([Fe/H]$\sim -1.5$; \citealp{gratton+05}), age ($t \sim 12.5$ Gyr;
e.g., \citealt{salaris+02, dotter+10, vandenberg+13}), HB morphology,
and HB mass distribution, NGC 6752 thus offers the ideal conditions to
further test the presence of a substantial population of slowly
cooling WDs in GCs with blue HB morphology.  Being closer to Earth
than M13, it also allows us to extend the study of this new class of
objects to fainter mangnitudes.

The paper is organized as follows. Section \ref{sec:data_reduction}
describes the adopted data reduction, calibration and
completeness-correction procedures. The analysis of the WD cooling
sequence and LF is presented in Section \ref{sec:resu}, which is
followed by the discussion of the results and conclusions in
Section~\ref{sec:conclu}.

\begin{table*}
\centering
\caption{Main physical parameters of NGC 6752.}
\label{table01}
\begin{tabular}{llll}
\hline \hline
Parameter		&			& &   Reference 					\\ \hline
%\multicolumn{3}{c}{NGC 6752} \\
$E(B-V)$		& $0.046\pm0.005$	& & \citealp{gratton+05}  \\
				& $0.040$	& & \citealp{harris+96}(2010 edition)\\
$ [Fe/H]$		& $-1.48\pm0.01$	& & \citealp{gratton+05}  \\
		 		& $-1.54$	& & \citealp{harris+96}(2010 edition)\\
$(m-M)_0$		&	13.13		& & \citealp{ferraro+99}\\
Age				& $12.5\pm0.75$	& Gyr & \citealp{dotter+10}	 \\
$\log{\nu_0}$ 	     & 5.04  			& L$_\odot$/pc$^3$ & \citealp{harris+96}(2010 edition)\\
$\log{t_{rc}}$ 	& 6.88  			& yr & \citealp{harris+96}(2010 edition)\\
\hline
\end{tabular}
\label{tab1}
\end{table*}

%%%%%%%%%%%%%%%%%%%%%%%%%%%%%%%%%%%%%%%%%%%%%%%%%%%%%%%%%%%
%\begin{table}
%\centering  
%\caption{Observations log of NGC 6752.}
  %The data set of long exposure frame are focus on white dwarf cooling
  %sequence (the main data sets), while short exposure frame are used
  %to focus on bright branch, e.g. HB and AGB.}
%\begin{tabular}{c|cc|cc}
%\hline \hline
%    			&	\multicolumn{2}{c}{Long exposure frames} & \multicolumn{2}{c}{{\textcolor{red} {Short exposure frames}}} \\ 
%   Filter		&  F275W 	 			& F336W 		   	& {\textcolor{red} {F225W}}      		&   {\textcolor{red} {F390W}}  			\\ 
%   Program ID 	&  GO-12311				& GO-11729			& GO-11904			& GO-11729			\\
%\hline
%$n\times t_{\rm exp}$&	$12\times 369$s  	&  $2\times 500$s 	& $6\times 120$s 	& $1\times 10$s 		\\
%       			&	 					&  $1\times 30$s  	&        		   	&  $2\times 2$s 		\\
%\hline
%\end{tabular}
%\label{table02}
%\end{table}

%% \begin{table}
%% \centering  
%% \caption{Observations log of NGC 6752.}
%% \begin{tabular}{c|cc}
%% \hline \hline
%%     			&	\multicolumn{2}{c}{Long exposure frames}   \\ 
%%    Filter		&  F275W 	 			& F336W 		   	     \\ 
%%    Program ID 	&  GO-12311				& GO-11729		\\
%% \hline
%% $n\times t_{\rm exp}$&	$12\times 369$s  	&  $2\times 500$s 		\\
%%        			&	 					&  $1\times 30$s   	\\
%% \hline
%% \end{tabular}
%% \label{table02}
%% \end{table}

%%%%%%%%%%%%%%%%%%%%%%%%%%%%%%%%%%%%%%%%%%%%%%%%%%%%%%%%%%%%%%%%%%%%%%%
\section{Data reduction}
\label{sec:data_reduction}
The high-resolution photometric data-set of NGC 6752 used in this work
has been acquired in the near-ultraviolet (near-UV) with the
\emph{HST} under GO-12311 (PI: Piotto) and GO-11729 (PI: Holtzman),
using the UVIS channel of the Wide Field Camera 3 (WFC3). It is
composed of 12 long exposures (369 s each) in the F275W filter and 3
images in the F336W ($2\times 500$ s, and $1\times 30$ s).

The photometric analysis was carried out on the $\_flc$ images (which
are the UVIS calibrated frames, also corrected for charge transfer
efficiency), after applying the Pixel Area Map (PAM) correction. We
used DAOPHOT IV \citep{stetson+87} to follow the so-called
\lq{UV-route\rq}, which consists in first searching for stellar
sources in the near-UV images, then force-fit the source detection at
the same positions of the UV-selected stars in longer wavelength
images \citep{ferraro+97b, ferraro+01, ferraro+03, raso+17,
  dalessandro+18a, dalessandro+18b, cadelano+19, cadelano+20a}. This
procedure allows the optimal recovering of blue and faint objects,
like WDs, because the crowding effect generated by giants and main
sequence turn-off (MS-TO) stars (which get brighter with increasing
wavelengths) is strongly mitigated in near-UV images of old stellar
populations (as Galactic GCs).

At first we selected $\sim 250$ bright and unsaturated stars,
relatively uniformly distributed in the sampled field of view, to
determine the Point Spread Function (PSF) function of each
exposure. The resulting model was then applied to all the sources
detected above 5$\sigma$ from the background, and the stars found at
least in half of the UV images were combined to create a
\textit{master list}.  The photometric fit was then forced in all the
other frames at the positions corresponding to the master list stars,
by using DAOPHOT/ALLFRAME \citep{stetson+94}.  Finally, the magnitudes
estimated for each star in different images of the same filter were
homogenized, and their weighted mean and standard deviation have been
adopted as the star's magnitude and its related photometric error.
The instrumental magnitudes were calibrated to the VEGAMAG system and
the instrumental coordinates, once corrected for geometric distortions
\citep{bellini+11}, have been put onto the International Celestial
Reference System by cross-correlation with the catalog obtained from
the \emph{HST} UV Globular Cluster Survey \citep{piotto+15}.  A
selection in sharpness was applied to remove non-stellar objects
(background galaxies) mainly affecting the faintest end of the sample.
All stars with sharpness parameter larger than 0.2 have been
removed. An accurate additional visual inspection of the images was
then necessary to decontaminate the sample from spurious detections
and artefacts in the regions sourrounding heavily saturated stars
\citep{annunziatella+13}.

%%===================================================================
\subsection{Color-magnitude diagram and comparison with M13}
\label{sec:cmd}
\begin{figure}
\centering
\includegraphics[width=12cm]{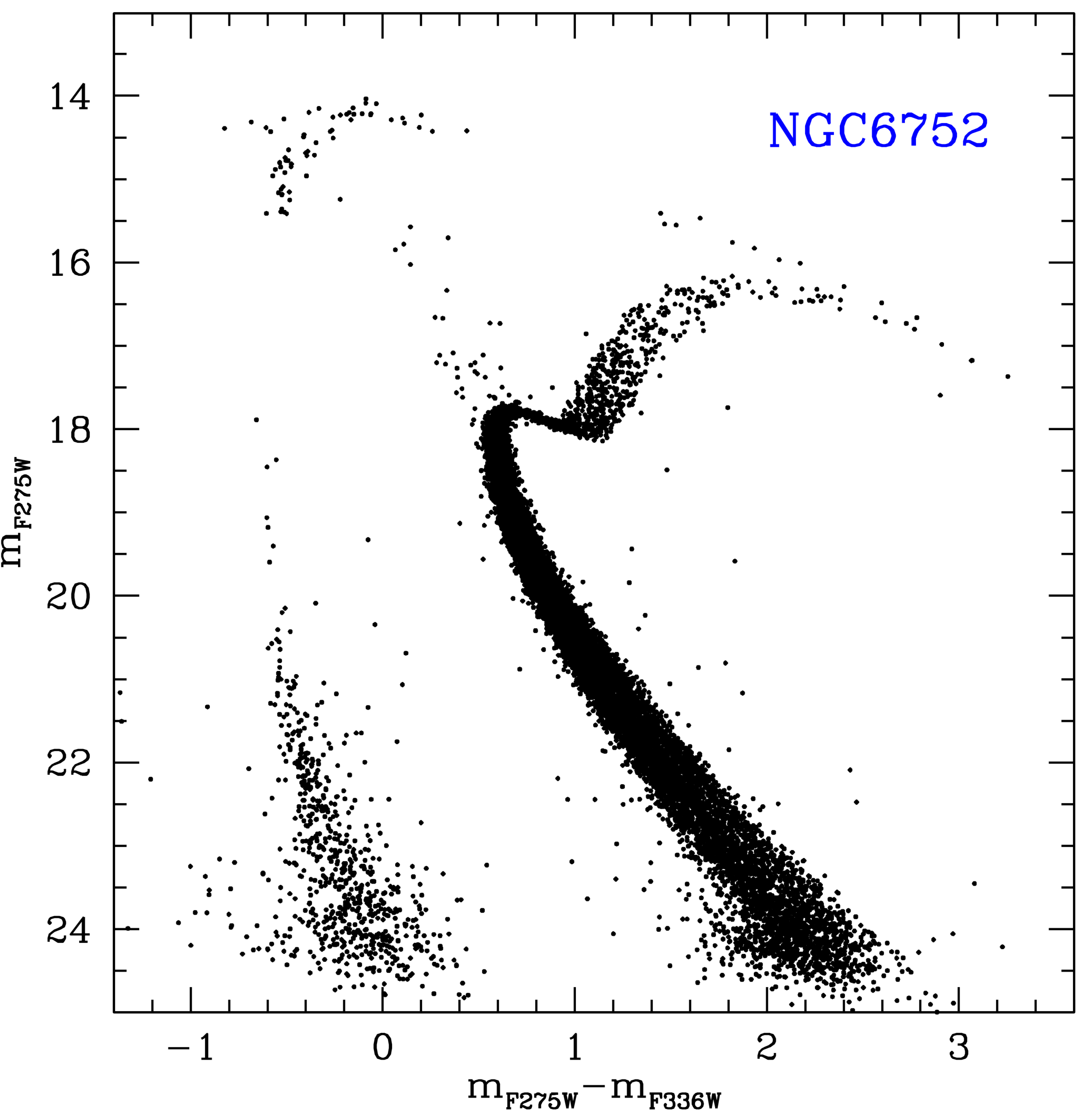}
\caption{Near-UV color-magnitude diagram (CMD) of NGC 6752.}
\label{fig:cmd}
\end{figure}

The final $(m_{\rm F275W}, m_{\rm F275W}-m_{\rm F336W})$ CMD of NGC
6752 is shown in Fig. \ref{fig:cmd}. It spans approximately 11
magnitudes, providing us with a full view of the cluster stellar
population down to nearly $m_{\rm F275W}=25$, and with a clear definition of
all the evolutionary sequences.  As expected in the UV-band, the
luminosity of RGB stars is substantially decreased with respect to
optical wavelengths, while hot HB stars are the brightest objects.
Also a sparse population of blue straggler stars is clearly visible
(see \citealt{sabbi+04}), covering a magnitude range comparable to
that spanned by the RGB.  Notably, the bright portion of the WD
cooling sequence appears well defined and populated, reaching
luminosities comparable to those of MS-TO stars, and covering $\sim 6$
magnitudes in the CMD.

\begin{figure}
\centering
\includegraphics[width=12cm]{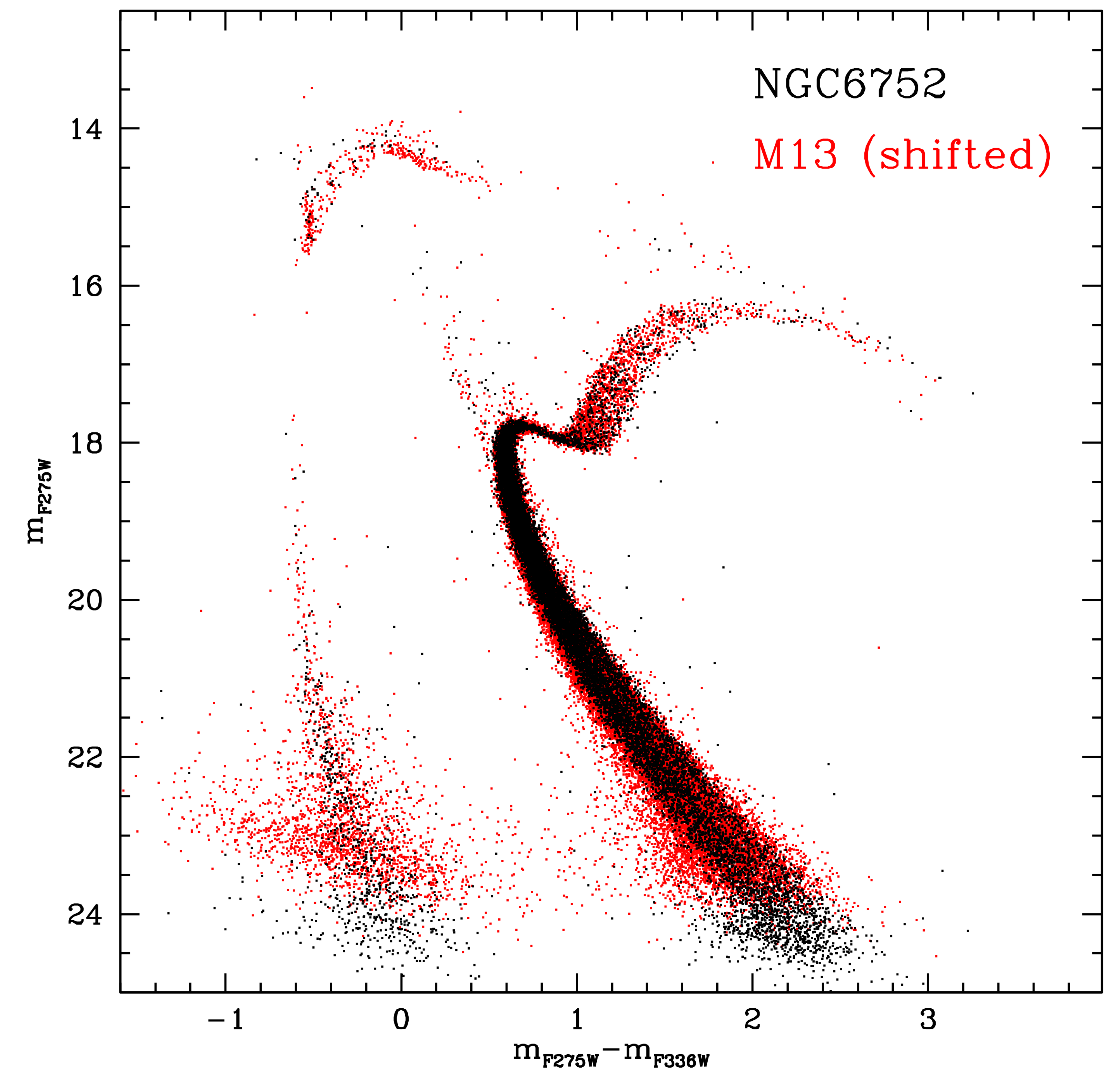}
\caption{Observed (Co-added $m_{\rm F275W}, m_{\rm F275W}-m_{\rm
    F336W}$ CMDs of NGC 6752 (black dots) and M13 (red dots, from
  C21). A magnitude shift $\Delta m_{F275W}=-1.04$ and a color shit
  $\Delta(m_{F275W}-m_{F336W})=0.04$ have been applied to the CMD of
  M13 to match that of NGC 6752.}
\label{fig:cmd_comb}
\end{figure}

The comparison with Fig. 1 of C21 reveals that the overall morphology
of the CMD evolutionary sequences in NGC 6752 is very similar to that
observed in M13.  Indeed, the MS-TO and the subgiant branch can be
adopted as optimal reference sequences to determine the shifts in
magnitude and color needed to superpose one CMD on the other.  We
found that a magnitude shift $\Delta m_{\rm F275W}=-1.04 \pm 0.02$ and
a small color shift $\Delta (m_{\rm F275W}-m_{\rm F336W})=+0.04 \pm
0.01$ are required to move the CMD of M13 onto that of NGC 6752. 
  The measured differences are compatible with the shorter distance
  ($\sim 3$ kpc closer, \citealt{ferraro+99}) and the larger reddening
  ($\delta E(B-V)=0.02-0.03$, see Table 1 and \citealt{harris+96}) of
  NGC6752 with respect to M13. Indeed the superposition provides an
impressive match of all the evolutionary sequences, including the
cooling sequence (see Fig.  \ref{fig:cmd_comb}). In particular we
notice that, although being less populated, the HB of NGC 6752 shares
the same morphology observed in M13: HB stars are distributed over the
same color range, $-0.6<(m_{\rm F275W}-m_{\rm F336W}) < 0.4$, with no
redder HB stars being detected in both clusters. Also the WD cooling
sequences are well superposed, with that in NGC 6752 reaching fainter
magnitudes, consistently with its shorter distance from Earth.

%%===================================================================
\subsection{Artificial star tests}
\label{sec:artif_stars}
To perform a quantitative study of the WD population of NGC 6752 in
terms of star counts, it is necessary to take into account the
photometric completeness of the WD cooling sequence at different
magnitudes and different distances from the cluster centre.  This is
particularly important in dynamically evolved stellar systems like NGC
6752, which is classified as a post core collapse cluster
\citep{harris+96}. Indeed, the innermost portion ($r<10\arcsec$) of
the stellar density profile shows a clear excess with respect to the
\citet{king+66} model that properly fits the external regions
\citep{ferraro+03}, consistently with its post core collapse
classification.\footnote{The advanced stage of dynamical evolution of
this system is further certified by the high radial concentration of
its blue straggler stars (see \citealt{lanzoni+16,ferraro+18}), which
is a powerful indicator of dynamical age (the “dynamical clock”, see
\citealp{ferraro+12,ferraro+18,ferraro+19,ferraro+20}).}

To estimate the photometric completeness, we performed artificial
stars experiments following the prescriptions described in detail in
\citet[][see also
  \citealt{dalessandro+15,cadelano+20b}]{bellazzini+02}, that we
summarize quickly here.  As first step, we created a list of
artificial stars with an input F275W magnitude sampling the observed
extension of the WD cooling sequence. Then, each of these stars was
assigned a F336W magnitude obtained by interpolating along the mean
ridge line of the WD cooling sequence in the $(m_{\rm F275W}, m_{\rm
  F275W}-m_{\rm F336W})$ CMD.  The artificial stars thus generated
were then added to the real images by using the DAOPHOT/ADDSTAR
software, and the entire photometric analysis was repeated following
exactly the same steps described in Section \ref{sec:data_reduction}.
To avoid artificial crowding, the added stars were placed into the
frames in a regular grid of 23$\times$23 pixels (corresponding to
about fifteen times the FWHM of stellar sources), each cell containing
only one artificial star during each run.
%We performed exactly the same photometric analysis as described in
%Section \ref{sec:data_reduction}.
The procedure was iterated several times and, at the end, more than
100,000 artificial stars were simulated in the entire field of view.

The ratio between the number of artificial stars recovered at the end
of the photometric analysis (number of output stars, $N_o$) and the
number of stars that were actually simulated (number of input stars,
$N_i$) defines the completeness parameter $\Phi = N_o/N_i$.  As it is
well known, the value of $\Phi$ strongly depends on both crowding
(hence, the distance from the cluster center) and stellar luminosity,
becoming increasingly smaller in cluster regions of larger density and
at fainter magnitudes. We thus split the sample of simulated stars into
bins of cluster-centric distances (stepped by $5\arcsec$) and F275W
magnitudes (from $\sim 18$, to $\sim 25$, in steps of 0.5~mag) and,
for each cell of this grid, we counted the number of input and
recovered stars, calculating the corresponding value of $\Phi$. The
sizes of the radial and magnitude bins were chosen to secure enough
statistics while keeping as high as possible both the spatial
resolution, and the sensitivity of the completeness curve to changes
in the stellar luminosity. The uncertainties assigned to each
completeness value ($\sigma_\Phi$) were then computed by propagating
the Poissonian errors, and typically are on the order of 0.05.  As
coordinates of the center of gravity we adopted the values quoted in
\citet{ferraro+03}: $C_{\rm grav}$ is located at $\alpha_{\rm J2000} =
19^{\rm h}\, 10^{\rm m}\, 52\fs04$, $\delta_{J2000} = -59\arcdeg\,
59\arcmin\, 04\farcs64$ with a typical $1\sigma$ uncertainty of
$0\farcs 5$ in both $\alpha_{\rm J2000}$ and $\delta_{J2000}$.The $C_{\rm grav}$ 
has been computed by averaging the coordinates of all stars
detected in the central portion of the cluster (see \citealt{miocchi+13,lanzoni+07,lanzoni+10,lanzoni+19})

\begin{figure}
\centering \includegraphics[width=10cm]{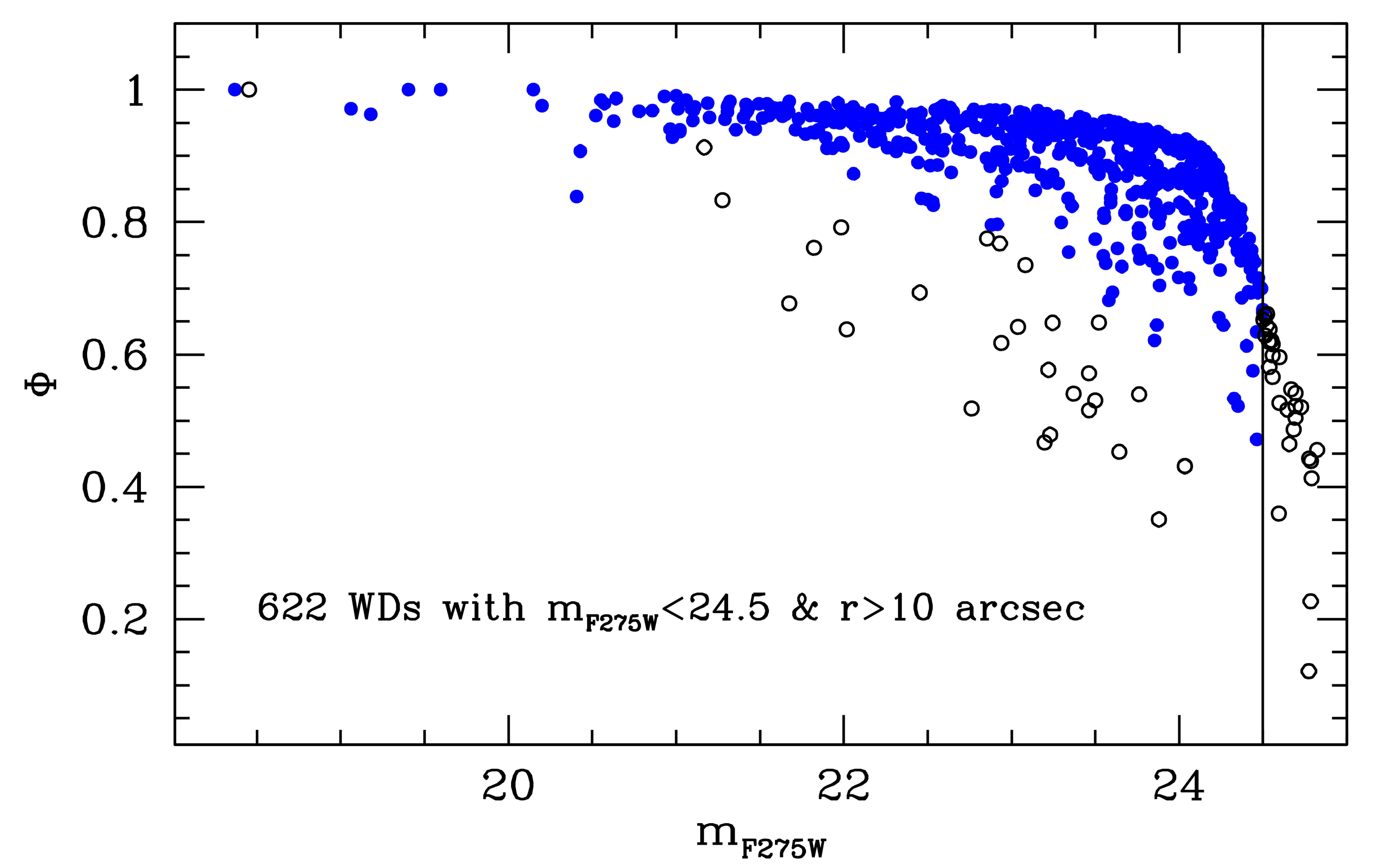}
\caption{Distribution of the WD completeness parameter $\Phi$, as a
  function of the F275W magnitude. The 622 selected WDs are plotted
  as blue circles: they all have $\Phi>0.45$, while those fainter
  than $m_{\rm F275W}=24.5$, or located at distances smaller than
  $10\arcsec$ from the cluster center (empty circles) have been excluded from
  the analysis, to avoid the risk of inappropriate completeness
  corrections.}
\label{fig:compl}
\end{figure}

The construction of such a completeness grid allowed us to assign the
appropriate $\Phi$ value to each observed WD, with a given F275W and
F336W magnitude, located at any distance from the cluster center. The
behavior of the completeness parameter as a function of magnitude is
shown in Fig.~\ref{fig:compl} for all the detected WDs.  By excluding
the overcrowded innermost region ($r<10\arcsec$) and the faintest end
($m_{\rm F275W}> 24.5$) of the cooling sequence (empty circles in the
figure), we obtain a well sampled population of 622 WDs with
completeness level larger than 45\% (blue circles).

%%%%%%%%%%%%%%%%%%%%%%%%%%%%%%%%%%%%%%%%%%%%%%%%%%%%%%%%%%%%%%%%%%%%%%%
\section{Analysis and results}
\label{sec:resu}

\subsection{Sample selection and WD LF}
\label{sec:LF}
\begin{figure}
\centering \includegraphics[width=7cm]{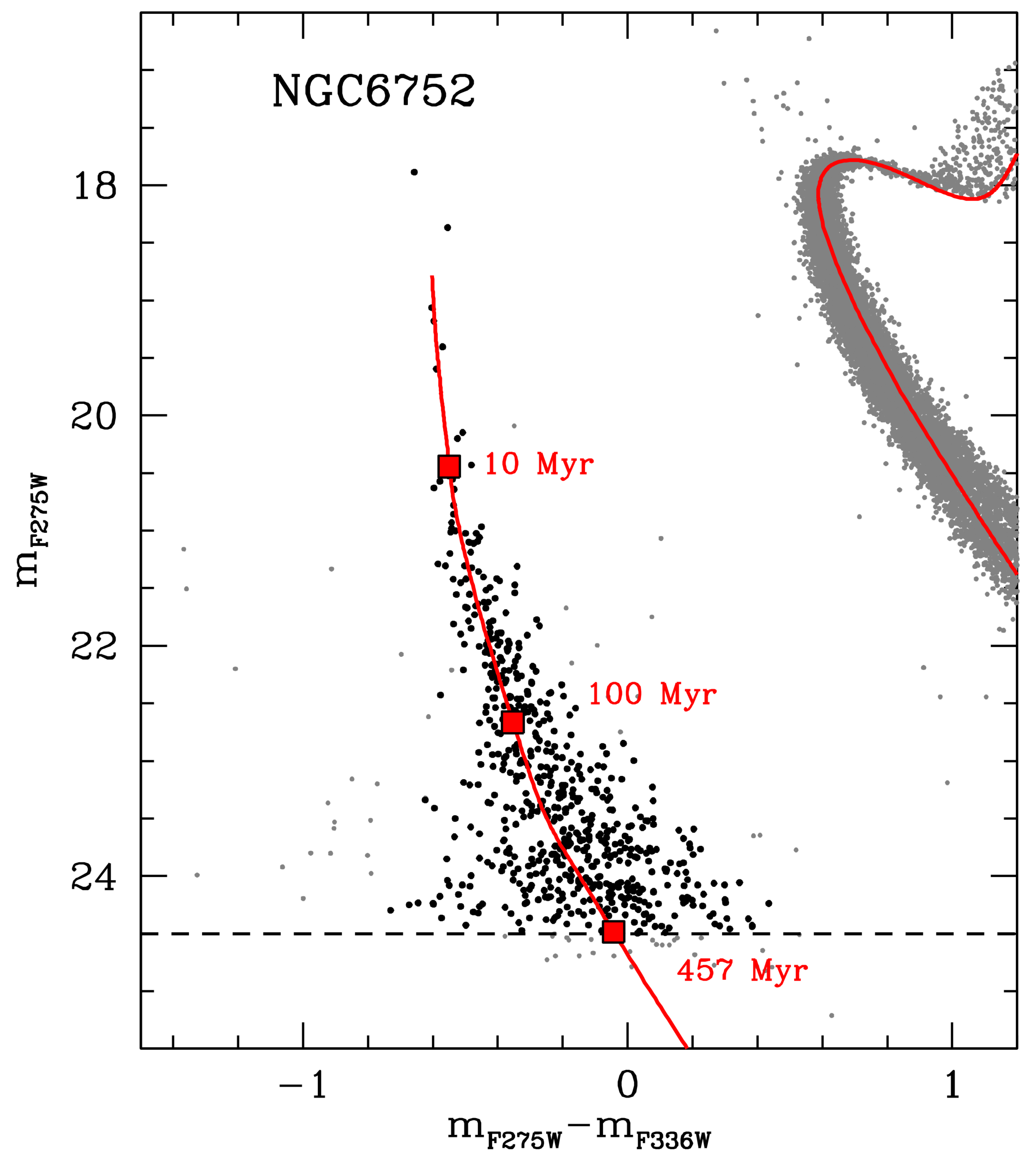}
\caption{CMD on NGC 6752 zoomed on the WD cooling sequence, with the
  WD samples selected for analysis denoted by black dots. The red
  lines are the same set of theoretical models used in C21 to
  reproduce the evolutionary sequences observed in M13: a 12.5 Gyr old
  isochrone from the BaSTI dataset \citep{pietrinferni+06} well
  fitting the cluster MS, and the cooling sequence of a $0.54 M_\odot$
  hydrogen atmosphere CO-WD from \citet{salaris+10}. The red squares
  flag three cooling ages along the cooling track. The dashed
  horizontal line marks the adopted magnitude limit of the analyzed WD
  sample, corresponding to a cooling age $t_{\rm cool}< 460$ Myr.}
\label{fig:wd_sample}
\end{figure}

The impressive match between the CMDs of NGC 6752 and M13 shown in
Fig.  \ref{fig:cmd_comb}) allows us to perform a direct comparison
among the properties of the two clusters, even using the same set of
isochrones and cooling tracks adopted in C21 for M13.  Indeed, Fig.
\ref{fig:wd_sample} shows the cooling track of a $0.54 M_\odot$
carbon-oxygen core WD with hydrogen atmosphere (DA-type) from
\citet{salaris+10} nicely reproducing the position of the observed WD
cooling sequence, and the 12.5~Gyr $\alpha$-enhanced isochrone with
metal abundance $Z=0.001$ and helium mass fraction $Y=0.246$ from the
BaSTI models (\citealp{pietrinferni+06}; see also \citealp{hidalgo+18,
  pietrinferni+21}) simultaneously reproducing the MS-TO region of the
cluster.

The well-defined cooling sequence in the CMD allowed us to select the
WD sample straightforwardly.  First, we excluded the objects located
more than 3$\sigma$ away from the mean ridge line of the WD cooling
sequence, with $\sigma$ being the photometric error at the
corresponding magnitude level.  Then, as discussed above, we
conservatively selected stars located at distances larger than
$10\arcsec$ from the cluster center and at magnitudes $m_{\rm
  F275W}\le 24.5$, to keep the completeness level above $45\%$.  The
adopted magnitude cut corresponds to a cooling time of $\sim 460$ Myr,
significantly longer than the case of M13, where the analysis was
limited to the first 100 Myr of cooling (see C21).  Thus, the
  cooling time sampled by the WD sequence in NGC 6752 is of the order
  of 0.5 Gyr. Although this is a significant range of cooling times,
  it corresponds to a negligible variation in terms of WD mass:
  indeed, the expected difference in mass between a WD that is
  currently starting the cooling process, and a WD with a cooling age
  of 0.5 Gyr is $0.001 M_{\odot}$ only.

\begin{figure}
\centering
\includegraphics[width=9.5cm]{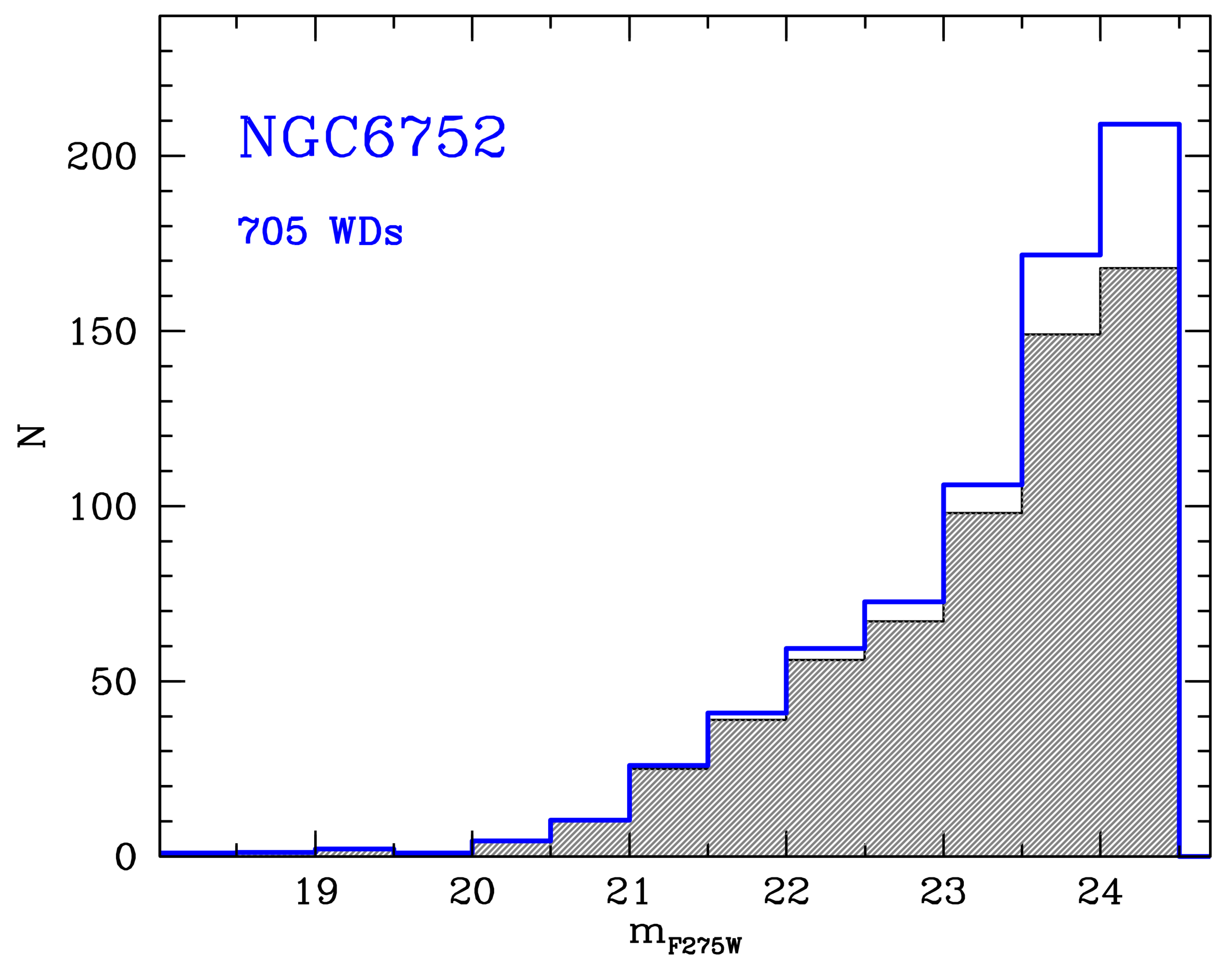}
\caption{Observed and completeness-corrected luminosity functions
  (shaded and blue histograms, respectively) obtained from the
  selected WD sample (black dots in Figure \ref{fig:wd_sample}).}
\label{fig:LFs}
\end{figure}

The adopted selection criteria provide a total of 622 WDs in NGC 6752.
The LFs computed in bins of 0.5 magnitudes are shown in Figure
\ref{fig:LFs} for both the observed and the completeness-corrected
samples (shaded and blue histograms, respectively). It can be seen
that the conservative criteria adopted for the sample selection
strongly limit the impact of incompleteness: the global correction to
the adopted samples is smaller than $15\%$, with the
completeness-corrected population of WDs counting 705 stars.

\subsection{Comparing the WD LF of NGC6752 and M13}
\label{sec:compare}

\begin{figure}
\centering
\includegraphics[width=9.5cm]{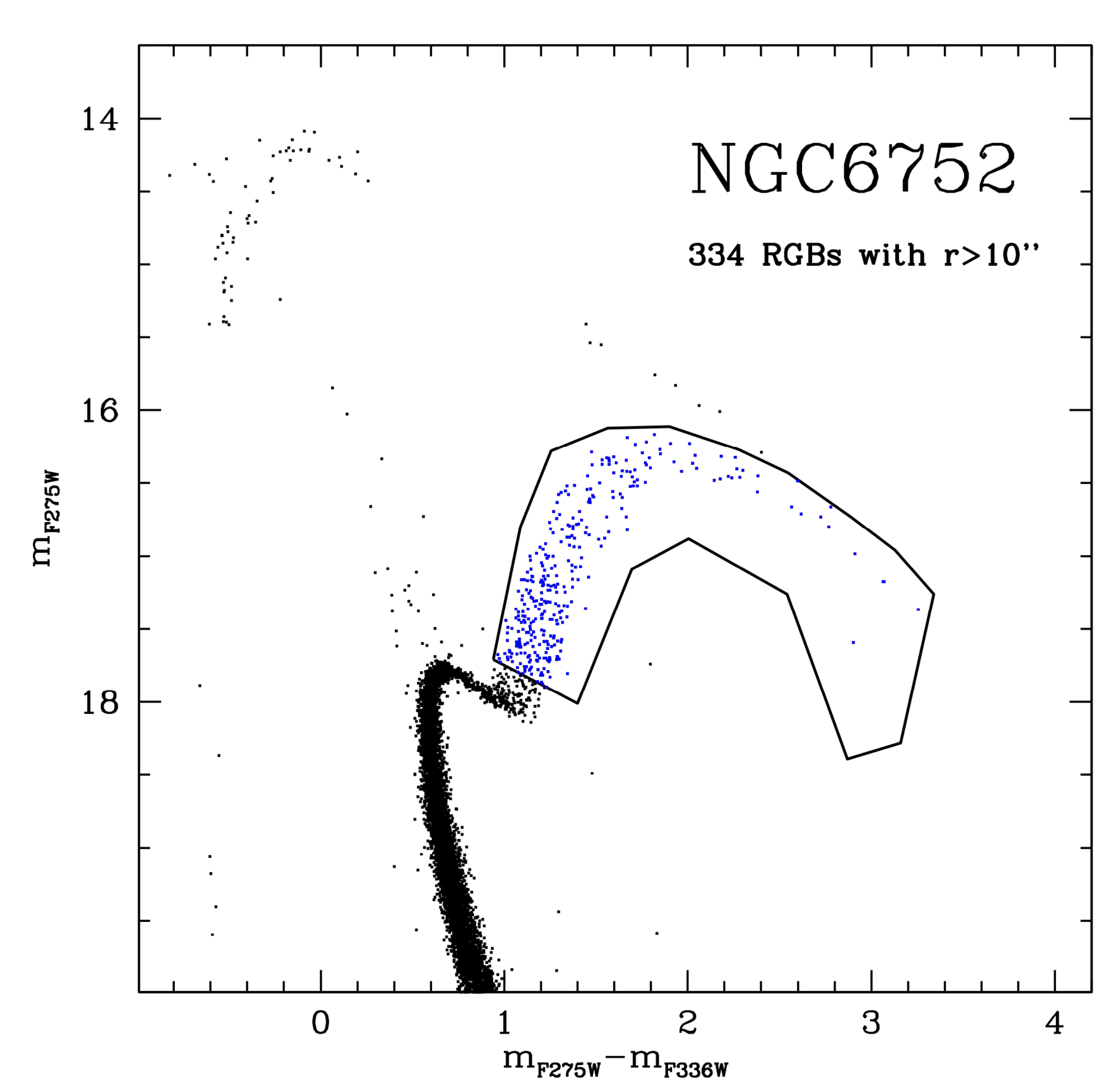}
\caption{Selection box (black line) of the RGB sample (blue dots) in
  NGC 6752.  The box is equal to that used in M13 (but for the color
  and magnitude shifts needed to align the two CMDs; see Figure
  \ref{fig:cmd_comb}). The number of RGB stars has been used as
  normalization factor of the WD samples to account for the different
  intrinsic richness of the two clusters.}
\label{fig:rgb}
\end{figure}

For a proper comparison between the WD LFs of NGC 6752 and M13, we
need to take into account the intrinsic mass of the two
clusters. According to the procedure adopted in C21, we used as
normalization the number of RGB stars. The stars are bright
(corresponding to high completeness levels at any distance from the
center), and for a given total stellar mass of the host population
their number depends only on the evolutionary speed, which is the same
in clusters with the same initial chemical composition and age (see,
e.g., \citealt{renzini+86}).  Taking advantage of the
excellent match obtained between the CMDs of the two clusters (see
Fig. \ref{fig:cmd_comb}), the RGB population was selected in the
observed CMD by using the same box adopted for M13, that samples the
entire RGB extension down to its base (see Figure \ref{fig:rgb}). Of
course, the RGB stars have been counted in the same cluster area
considered for the WD selection ($r>10\arcsec$), providing a total of
335 objects.

\begin{figure}
\centering
\includegraphics[width=9.5cm]{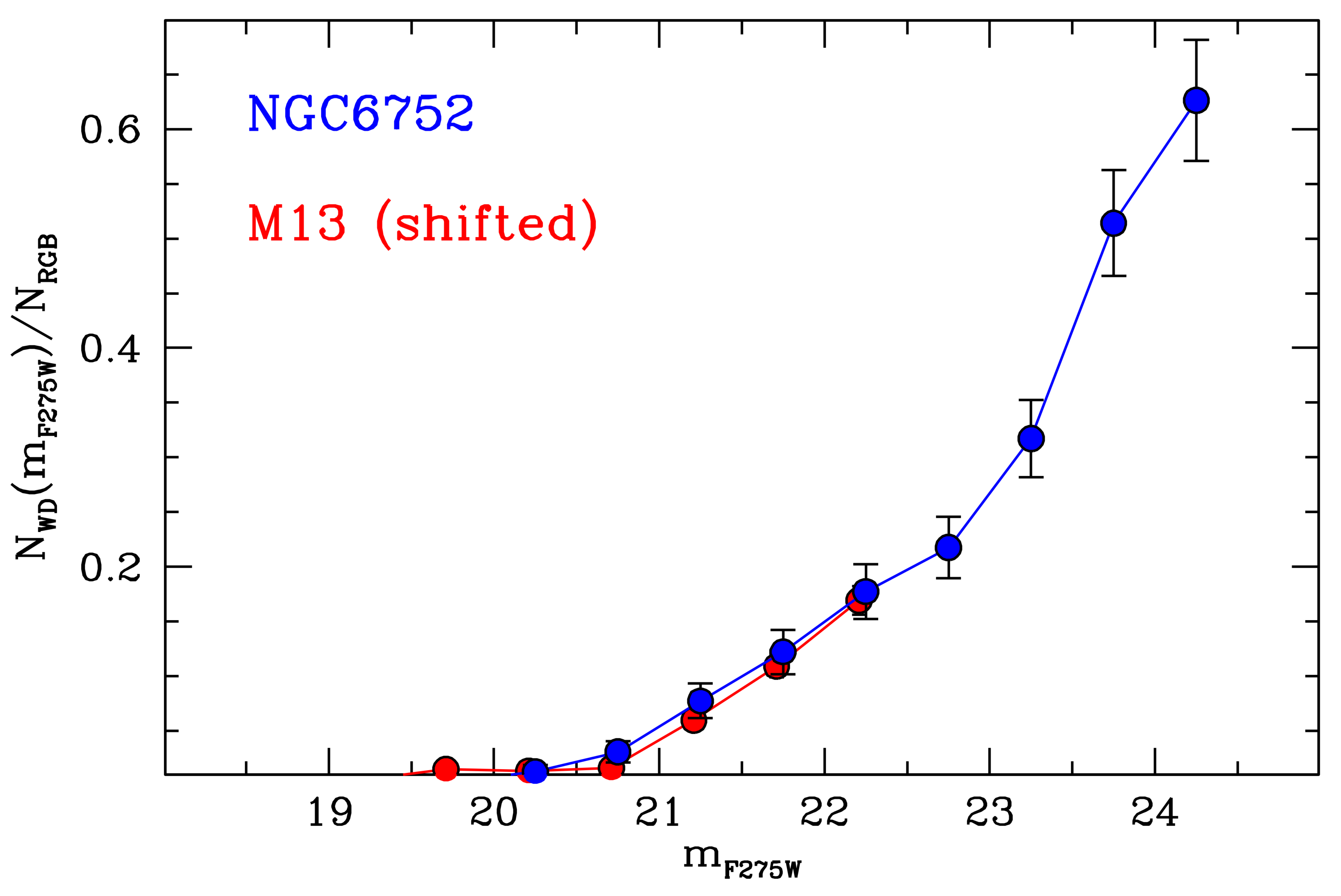}
\caption{Completeness-corrected WD differential LFs, normalized to the
  number of RGB reference stars, for NGC~6752 (blue circles) and M13
  (red circles). The same magnitude shift adopted in Fig.
  \ref{fig:cmd_comb}) has been applied to the LF of M13.}
\label{fig:lf_theory}
\end{figure}

The completeness-corrected number of WDs counted in the various
magnitude bins, divided by the total number of RGB stars thus
obtained, finally provided us with the WD LF normalized to the RGB
reference population, shown in Fig.~\ref{fig:lf_theory} (blue
circles). The corresponding normalized WD LF of M13 (from C21) is
plotted in the same figure in red for comparison. Although the WD LF
of NGC 6752 reaches much fainter magnitudes than that of M13, the
match between the results in the two systems is really impressive,
with the portions in common being almost indistinguishable. This
provides strong evidence that the two clusters share the same type of
WD population, suggesting that also NGC 6752 hosts a relevant fraction
of slowly cooling WDs, as expected from its HB morphology.

This expectation is also consistent with the conclusion by \citet{cassisi+14}
that the mass distribution along the HB in the two clusters is
essentially the same, with the large majority of HB stars having
masses smaller than $\sim 0.56 M_\odot$. According to theoretical
models (BaSTI and \citealt{bono+20}), at the
considered metallicity ($Z=0.001$), these stars do not experience the
thermal pulse AGB stage during which the third dredge-up occurs. We can thus assume that, as in M13,
approximately $70\%$ of the WD progenitors in NGC 6752 end their
evolution with hydrogen envelopes thick enough (a few $10^{-4}
M_\odot$) to guarantee stable H-burning during the subsequent WD
cooling stage. The same proportion between standard and slowly cooling
WDs ($\sim 30$\% and 70\%, respectively) is therefore expected in the
two clusters.

\begin{figure}
\centering
\includegraphics[width=12cm]{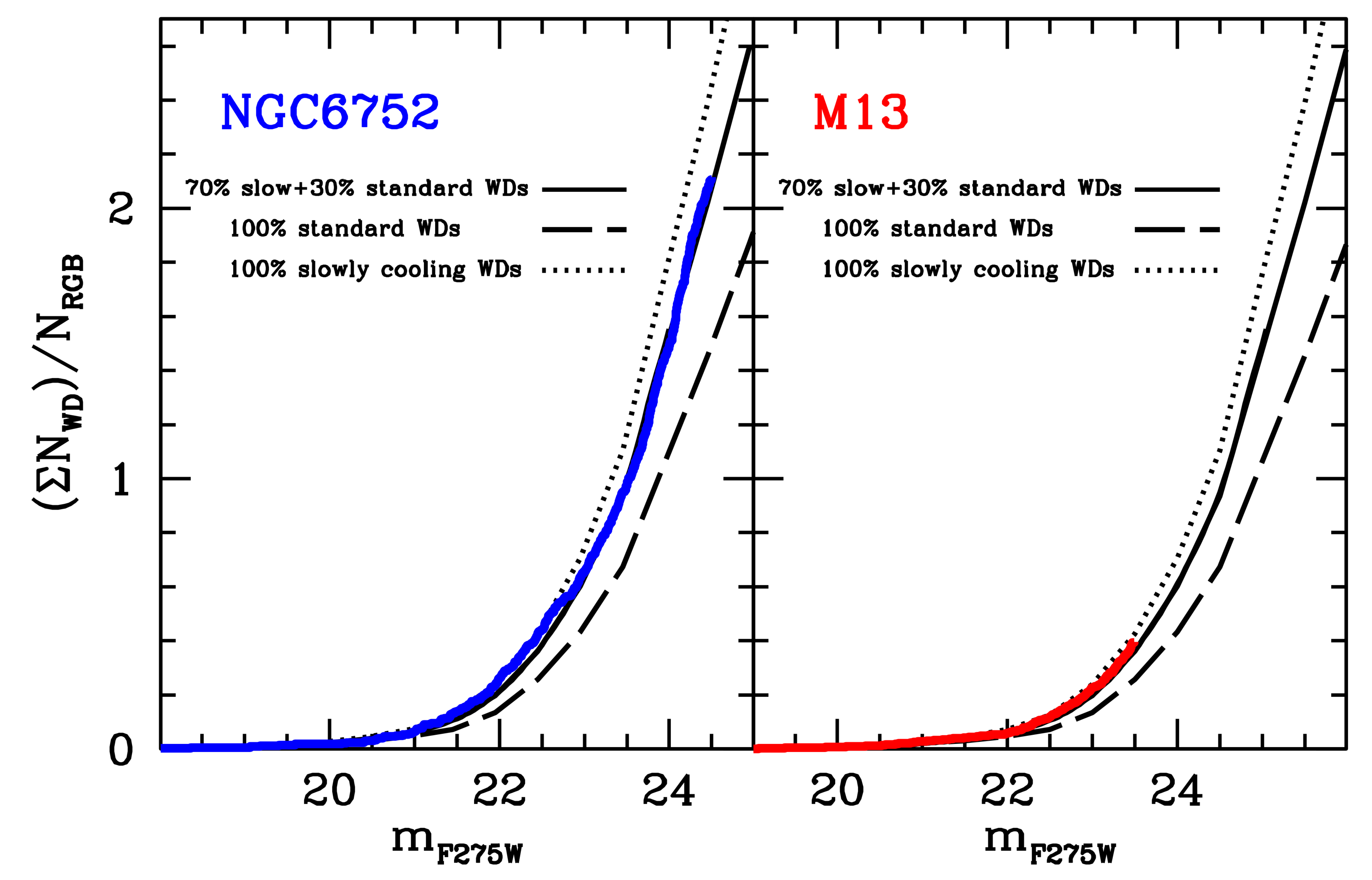}
\caption{Left panel: WD cumulative LF of NGC 6752, corrected for
  completeness and normalized to the number of RGB stars (thick blue
  line), compared to the results of Monte Carlo simulations including
  100\% standard WDs (dashed line), 100\% slowly cooling WDs
    (dotted line) and a combination of 30\% standard WDs and 70\%
  slowly cooling WDs (black solid line). Right panel: the same for M13
  (red thick line, from C21). }
\label{fig6752_m13}
\end{figure}

To quantitatively test this prediction, the left panel of
Fig.~\ref{fig6752_m13} compares the completeness-corrected cumulative
LF, normalized to the number of RGB stars, of NGC 6752 WDs (thick blue
line), with the results of Monte Carlo simulations of the entire
evolutionary path from the RGB, to the HB and then the WD stages,
specifically computed by adopting the HB mass distributions determined
for M3 and M13 (see C21).  The simulation including 100\% of standard
WDs (as observed in M3) corresponds to the dashed line in the figure,
while the run including a combination of 30\% standard WDs and 70\%
slowly cooling WDs is marked with the black solid line.  As expected,
the WD LF of NGC 6752 is in impressive agreement with the latter,
demonstrating that, according to the cluster HB morphology, the WD
population is dominated by a relevant fraction of slowly cooling
objects.  For the sake of comparison, the right panel of Figure
\ref{fig6752_m13} shows the normalized WD cumulative LF of M13
determined by C21, together with the same models plotted in the left
panel.  The faintest WDs sampled in NGC 6752 have luminosities of just
$\sim 10^{-2.7} L_\odot$, significantly fainter than those probed in
M13. Thus the results presented in this work provide further solid
support to the existence of slowly cooling WDs, and to the scenario
traced in C21 about their origin.
 
\begin{figure}
\centering
\includegraphics[width=9.5cm]{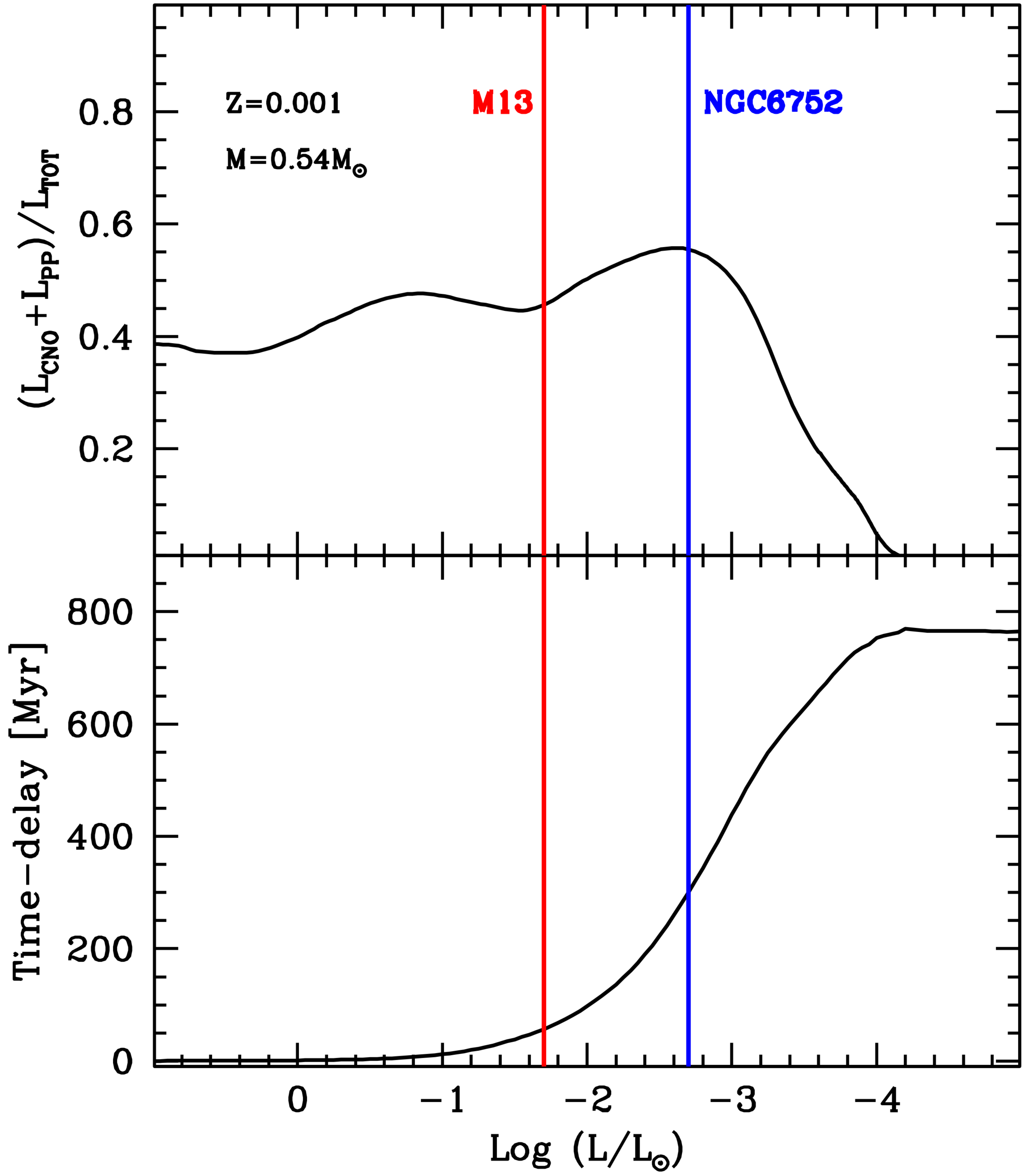}
\caption{Top panel: Contribution of stable H-burning (via PP and CNO
  chain) to the luminosity of a low metallicity (Z=0.001), low mass
  (0.54 $M_{\odot}$) WD as a function of the luminosity.  Bottom
  panel: Delay in the cooling time induced by stable H-burning, with
  respect to a model without burning. The two vertical lines mark the
  luminosity reached in NGC 6752 (present work) and in M13 (C21).}
\label{fig:delay}
\end{figure}

%%%%%%%%%%%%%%%%%%%%%%%%%%%%%%%%%%%%%%%%%%%%%%%%%%%%%%%%%%%%%%%%%%%%%%%
\section{Conclusions}
\label{sec:conclu}
The key points linking the presence of slowly cooling WDs in a cluster
to its HB morphology can be summarized as follows.
\begin{enumerate}
\item The energy support responsible for slowing down the cooling
  process is provided by stable H-burning in the WD external layers,
  which is made possible by the presence of a H-rich residual envelope
  containing enough mass (on the order of a few $10^{-4} M_\odot$;
  see, e.g., \citealt{renedo+10}). More standard WDs like those in M3
  have a H content below the threshold for H-burning ignition, and
  thus progressively cool in time with no sources of active energy
  production.
\item According to \citet{althaus+15}, the third dredge-up is the
  crucial event regulating the mass of the residual hydrogen in
  proto-WDs. In fact, during the third dredge-up carbon is carried
  into the convective envelope and, in turn, hydrogen is brought down
  inside the star, where it is burned. Thus, the occurrence (or lack
  thereof) of the third dredge-up affects the residual H envelope of
  the proto-WD when it reaches its cooling track, impacting its
  structure and cooling time.
\item Theoretical models (e.g., \citealt{bono+20})
  show that at $Z=0.001$, HB stars with masses smaller than $\sim 0.56
  M_\odot$ skip the thermal pulse AGB stage, where the third dredge-up
  occurs. This is because their envelope mass is so small that the
  alternate helium and hydrogen burning in shells surrounding the
  carbon-oxygen core cannot occur.  Completely or partially skipping
  the AGB phase guarantees the survival of a residual hydrogen
  envelope in these stars when they evolve off the HB toward the WD
  stage: Hence, they will evolve as WDs with stable H-burning and a
  significant delay of their cooling time.
\item From stellar evolution theory it is also well known that HB 
  stars with the above mentioned mass and metallicity populate the
  blue tail of the observed HBs.
\end{enumerate}
As a result of this chain of events, significant populations of slowly
cooling WDs are expected in globular clusters with well populated and extended blue 
HB morphologies.

By showing that NGC 6752, a Galactic GC with blue extended HB morphology,
hosts a large sample ($\sim 70\%$ of the total) of slowly cooling WDs,
this work not only provides new evidence for the existence of this class
of WDs, but also gives strong support to the picture above. Indeed,
the fact that the same mixture of standard and slow WDs is able to
reproduce the WD LFs observed in both NGC 6752 and M13, while 100\% of
standard WDs are needed for M3 (see C21), demonstrates the link
between the cluster HB morphology and the presence of these stars, and
provides empirical support to the proposed physical origin of these
objects (i.e., the fact that they skip the third dredge-up). In fact,
the only major difference between M3 and the other two systems is
the HB morphology: at odds with M3, NGC 6752 and M13 have a
(remarkably similar) blue-tail HB, which is populated by such low-mass
stars that they skip the third dredge-up and evolve in slowly cooling
WDs.

While the WD cooling sequence in M13 samples luminosities down to
$log(L/L_\odot)\sim -1.7$ (see C21), in NGC 6752 we reached
luminosities one order of magnitude fainter, corresponding to much
longer delay times with respect to the standard predictions for a pure
cooling process. This is shown in Fig.~\ref{fig:delay}, where
the cooling time delay induced by stable H-burning, with respect to
standard models of WDs with no burning is plotted as a
function of the stellar luminosity. While the magnitude limit reached
in M13 corresponds to a delay of $\sim 60$ Myr, the limit achieved in
NGC 6752 implies a delay time of $\sim 300$ Myr, meaning that most of
the faintest WDs in our adopted photometry have cooling ages of $\sim 760$ Myr,
instead of $\sim 460$ Myr as standard models would suggest
(see Figure \ref{fig:wd_sample}).

The figure also shows that the contribution to the energy budget
provided by the stable burning in the thin H-rich layer is essentially
restricted to $\log(L/L_{\odot}) > -$4 for a low mass WD (0.54
$M_{\odot}$) at intermediate metallicity (Z=0.001). According to the
models, this luminosity corresponds to a cooling age of approximately
3.5 Gyr, and during this time the accumulated delay in the cooling
process can be as large as 0.8 Gyr . An exhaustive exploration of the
entire H-burning phase down to magnitudes corresponding to at least
$\log(L/L_{\odot})\sim -$4 is thus required, to obtain an empirical
measurement of the total amount of cooling delay induced by this
mechanism.

 It is worth emphasising that the predictions shown in Figure 9
  refer to a single-mass WD, whilst the modelling of the faintest end
  of the WD LF is much more complex. In fact, the bright portion of
  the cooling sequence can be well approximated by a single mass
  ($\sim 0.54 M_\odot$) cooling track, but at the faintest end the
  contribution of WDs with larger masses must be taken into account to
  properly characterize the impact of the H-burning on the WD LF. {\it
    In the case of NGC 6752, which is the portion of the WD cooling
    sequence that is expected to be affected by the presence of slow
    WDs?} As discussed above, slowly cooling WDs are the progeny of
  the (bluest) HB stars with mass below $0.56 M_\odot$, that skip the
  AGB thermal pulses. In 12-13 Gyr old clusters, the initial mass of
  stars on the RGB scales as $dM/dt=-0.02 M_{\odot}/Gyr$. Hence, given
  the current RGB mass in NGC 6752 ($\sim0.8 M_{\odot}$), we expect
  that about 3 Gyr ago RGB stars had a mass around $0.86
  M_\odot$. Under the reasonable assumption that the amount of mass
  lost along the RGB is approximately constant for low-mass stars of
  similar initial masses, this implies that about 3 Gyr ago (and even
  earlier in the cluster life) all the core He-burning stars had a
  mass above $0.56 M_{\odot}$ and thus have all evolved through the
  thermal pulses along the AGB, producing 'normal' WDs. Hence, all the
  WDs with cooling age larger than 3 Gyr (which corresponds to
  $m_{F275W}\sim 31$) are expected to be normal WDs, and slowly
  cooling WDs are predicted to affect only the portion of the WD
  cooling sequence brighter than $m_{F275W}\sim 31$, with an
  importance that progressively decreases at fainter magnitudes down
  to this limit.

While the results presented here and in C21 provide a solid
  demonstration of the existence of slowly cooling WDs at intermediate
  metallicities ($Z =0.001$), the extension of this investigation to
  fainter magnitudes and other metallicity regimes is now necessary to
  fully verify the theoretical predictions and provide an empirical
  measure of how common this phenomenon is in stellar systems. This is
  of paramount importance for the correct use of the WD cooling
  sequences as chronometers to measure cosmic ages.

%%%%%%%%%%%%%%%%%%%%%%%%%%%%%%%%%%%%%%%%%%%%%%%%%%%%%%%

\acknowledgments This research is part of the project {\it Cosmic-Lab}
at the Physics and Astronomy Department of the University of Bologna
(http://www.cosmic-lab.eu/Cosmic-Lab/Home.html). The research has been
funded by project {\it Light-on-Dark}, granted by the Italian MIUR
through contract PRIN-2017K7REXT (PI: Ferraro). JXC acknowledges the
support from China Scholarship Council (CSC). MS acknowledges support
from The Science and Technology Facilities Council Consolidated Grant
ST/V00087X/1.

%%%%%%%%%%%%%%%%%%%%%%%%%%%%%%%%%%%%%%%%%%%%%%%%%%%%
%%%%%%%%%%%%%%%%%%%%%%%%%%%%%%%%%%%%%%%%%%%%%%%%%%%%
\vspace{5mm}

\facilities{HST(WFC3)}
\software{DAOPHOT\citep{stetson+87}, ALLFRAME\citep{stetson+94} }

%\appendix
%\section{Appendix information}

\vskip12pt
\newpage
%\bibliography{}{}

%\bibliography{ngc6752I_Chen_v1}{}
%\bibliographystyle{aasjournal}

\end{document}